\documentclass[aps,prd,reprint,superscriptaddress,nofootinbib,longbibliography]{revtex4}%
\usepackage{graphicx}
\usepackage{bm,latexsym,amsmath,amssymb,amsfonts,mathrsfs}
\usepackage{color}
\input{colordvi.tex}

\usepackage{url}
\usepackage{ulem}
\usepackage{hyperref}
\hypersetup{
 colorlinks=true,
  citecolor=cyan,
 }
\allowdisplaybreaks[1]
 \usepackage{xcolor}
\usepackage{float}
\usepackage[hang,small,bf]{caption}
\usepackage[margin=0pt]{subcaption}
\newcommand{\md}{\mathrm{d}}
\usepackage{natbib}
\usepackage{hyperref}
\hypersetup{
    colorlinks=true,
    linkcolor=blue,
    filecolor=blue,      
    urlcolor=blue,
    citecolor=blue,
} 
\usepackage[compat=1.1.0]{tikz-feynman}

\begin{document}

\title{Constraining the Gravitational Wave Speed in the Early Universe via Gravitational Cherenkov Radiation}
\author{Paola C. M. Delgado}
\email[Email: ]{delgado@fzu.cz}
\affiliation{CEICO, FZU – Institute of Physics of the Czech Academy of Sciences, Na Slovance 1999/2, 182 00 Prague, Czech Republic
}
\author{Alexander Ganz}
\email[Email: ]{alexander.ganz@itp.uni-hannover.de}
\affiliation{Institute for Theoretical Physics, Leibniz University Hannover,
Appelstraße 2, 30167 Hannover, Germany
}
\author{Chunshan Lin}
\email[Email: ]{chunshan.lin@uj.edu.pl}
\affiliation{Faculty of Physics, Astronomy and Applied Computer Science, Jagiellonian University, 30-348 Krakow, Poland
}
\author{Roxane Thériault}
\email[Email: ]{roxane.theriault@doctoral.uj.edu.pl}
\affiliation{Faculty of Physics, Astronomy and Applied Computer Science, Jagiellonian University, 30-348 Krakow, Poland
}

\begin{abstract}
Scalar particles traveling faster than a subluminal gravitational wave generate gravitons via gravitational Cherenkov radiation. In this paper, we investigate graviton production by the primordial plasma within the framework of modified gravity in the early Universe, generating a relic graviton background. We find that for the minimal model, where only the speed of  gravitational waves is modified and a standard model plasma minimally couples to gravity, the relic graviton background can be enhanced by several orders of magnitude, but still agrees with  the Big Bang Nucleosynthesis (BBN) bound in most cases. Moreover, we also find that for Horndeski theories, such as Galileon theory, the relic background produced by the thermalized scalar field can reach significant amplitudes, exceeding the BBN bound for a region of the parameter space. By requiring the relic graviton background to remain consistent with the BBN constraint, we derive limits on the gravitational wave speed at early times in these modified gravity theories.
\end{abstract}

\maketitle


\section{Introduction}\label{Sec:Introduction}
The observation of the binary neutron star merger GW170817 over the past decade \cite{LIGOScientific:2017vwq,LIGOScientific:2017zic,LIGOScientific:2017ync} imposed severe bounds on the speed of gravitational waves at late times in the LIGO-Virgo frequency range, namely $-3\times 10^{-15} \leq c_T - 1 \leq7\times 10^{-16}$. A decade earlier, the speed of high-frequency gravitational waves in 4D Lorentz-violating models was constrained through the observation of high-energy cosmic rays \cite{Moore:2001bv}. The latter would have generated gravitons via gravitational Cherenkov radiation if gravitational waves were subluminal and, therefore, their existence sets a bound on the gravitational wave speed, namely $c-c_T\leq 2\times 10^{-15}c$ for galactic-origin cosmic rays and $c-c_T\leq 2\times 10^{-19}c$ for extragalactic ones. 

While these constraints are stringent at late times, the early-time speed of gravitational waves remains less explored, and modified gravity theories that introduce $c_T\neq c$ remain, in principle, allowed. However, relativistic particles traveling faster than subluminal gravitational waves produce gravitons through the gravitational Cherenkov effect, generating a relic graviton background that can persist to the present day. On the other hand, Big Bang Nucleosynthesis (BBN) sets a bound on the graviton contribution to the total energy density in the early Universe. This constraint is expressed as a change in the effective number of relativistic species, $\Delta N_{\rm eff}$, which is limited to $\Delta N_{\rm eff}\lesssim 0.2$ at $95\%$ confidence level \cite{Planck:2018vyg,Fields:2019pfx} and translated to the fractional energy density of the background today as $h_0^2\Omega_{\rm g} \lesssim 1.12\times 10^{-6}$ \cite{Caprini:2018mtu,Bernreuther:2022hdl}. 
Therefore, one can constrain the gravitational wave speed at early times by requiring that the BBN bound is satisfied. On the other hand, 
direct cavity experiments are still not competitive at these high frequencies  against the BBN bound \cite{Domcke:2022rgu,Berlin:2021txa,Aggarwal:2020olq}. 

Most cosmological models predict a background of gravitational waves produced by a Standard Model plasma in thermal equilibrium in the early universe \cite{Guzzetti:2016mkm,Ringwald:2020ist,Ghiglieri:2020mhm}. This background is of the order $h_0^2\Omega_{\rm GWB} \sim 10^{-10}$ for $T_{\rm max} \sim 10^{15}$GeV \cite{Ringwald:2020ist}. Interestingly, this amplitude can be enhanced by several orders of magnitude if we  lower the speed of gravitational waves so that the primordial plasma, which is represented by a thermalized canonical scalar field in our simplified scenario, can convert to the relic graviton background via the gravitational Cherenkov radiation effect.
Namely, we start by assuming a modification to $c_T$ only, without specifying the theory that generates such modification and with a minimally coupled scalar field representing the primordial plasma. We shall refer to it as the minimal scenario from now on. Then, we consider Horndeski theories and the particular case of Galileon theory, in which the relic graviton background overcomes the BBN bound for a region of the parameter space, imposing bounds on the speed of gravitational waves in the early Universe. 

The paper is organized as follows: Section \ref{Sec:Gravitational Cherenkov Interaction} introduces the gravitational Cherenkov radiation and the resulting squared transition amplitude. In Section \ref{Sec:Minimal Scenario}, we compute the interaction rate and the relic graviton background for the minimal scenario, while in Section \ref{Sec: Quartic Horndeski} we carry out the analysis for Horndeski theories, including the specific example of Galileon theory. Finally, Section \ref{Sec:Conclusion} presents a summary of our findings and discusses their broader implications.

\section{Gravitational Cherenkov Interaction}\label{Sec:Gravitational Cherenkov Interaction}

Cherenkov radiation occurs when a massive charged particle travels faster than the phase velocity of light in a medium, which causes it to emit a photon. Gravitational Cherenkov radiation works analogously to its electromagnetic counterpart, but in this case, a particle travels faster than the phase velocity of gravity and emits a graviton. We consider the emission of a graviton with four-momentum $K^\mu =(\omega, \vec k)$ by a scalar particle traveling faster than the graviton speed $c_T$ with four momentum $(P^\prime)^\mu = (E^\prime , \vec p^\prime)$. After emission, the particle has the momentum $P^\mu = (E, \vec p) $. The Feynman rule for a $X \rightarrow X + h$ process, where $X$ in this case is a scalar particle and $h$ is a graviton, is given in \cite{Kosteleck__2015}. 
Thus, the Cherenkov radiation matrix element squared is 
\begin{align}
    \vert {\cal M} \vert^2 = 16 \pi G \sum \vert \epsilon^{\mu\nu} P_\mu (P+ K)_\nu \vert^2, 
\end{align}
using that the sum over the polarizations can be expressed as
\begin{align}
    \sum_\sigma \epsilon^{\sigma}_{ij} \epsilon^{\star \sigma}_{mn} = \lambda_{im} \lambda_{jn} + \lambda_{in} \lambda_{jm} - \lambda_{ij} \lambda_{mn}, 
\end{align}
where $\lambda_{ij}= \delta_{ij} - k_i k_j/k^2$. Then, without loss of generality, we can set the coordinate system such that $\vec p \cdot \vec k = p k \cos \theta$, where $\theta$ is the angle between $\vec p$ and $\vec k$, $|\vec p|\equiv p$, and $|\vec k|\equiv k$.  This results in
\begin{align}\label{eq: M matrix minimal scenario}
    \vert {\cal M} \vert^2 = 16 \pi G \left(p^2 - \frac{(\vec p \cdot \vec k)^2}{k^2} \right)^2 = 16 \pi G p^4 \sin^4 \theta.
\end{align}

\section{Minimal Scenario}\label{Sec:Minimal Scenario}

To obtain the GW spectrum in the thermal plasma we follow the approach in \cite{Ghiglieri:2020mhm,Ghiglieri:2022rfp}. The Boltzmann equation for the polarization averaged phase-space distribution of the gravitons $f_{h}$ can be written as
\begin{align}
    \label{eq:Boltzmann_equation}
    \frac{{\rm d} f_h(t,k)}{{\rm d} t} = \Gamma \left( f_\phi - f_h \right) ,
\end{align}
where $f_\phi$ is the phase space distribution for the scalar field and $\Gamma$ is the production rate of the gravitons due to the Cherenkov radiation. We consider an initially negligible distribution of gravitons, $f_h\ll 1$, and a distribution $f_{\phi}$ of scalars in thermal equilibrium, which does not change significantly over time, i.e.
\begin{align}
    f_h(t,k) = \delta f_h(t,k), \qquad f_\phi(t,k) = n_B(t,k) + \delta f_\phi(t,k),
\end{align}
where $n_B(t,p) = 1/(e^{E/T}-1)$ is the Bose-Einstein distribution and the $\delta 's$ denote the sub-leading-order terms.
Therefore, we can neglect the term $-\Gamma f_{h}$ in \eqref{eq:Boltzmann_equation}.
Expanding the Boltzmann equation to first order we get
\begin{align}
    \frac{{\rm d} \delta f_h(t,k)}{{\rm d} t} \simeq \Gamma f_\phi =& \frac{1}{4 \omega} \int \frac{\md^3p^\prime}{(2\pi)^3 2 E^\prime} \frac{\md^3 p}{(2\pi)^3 2 E} (2\pi)^4 \delta^{(4)}(P^\prime - P - K) \vert {\cal M} \vert^2 n_B(t, \vec p^\prime) (1 + n_B(t,\vec p) ), \\
    =& \frac{1}{4\omega} \int \frac{\md^3 p}{(2\pi)^3 2 E} \frac{2\pi}{2 E^\prime}  \delta(E^\prime - E - \omega) n_B(t, \vec p + \vec k) (1+n_B(t, \vec p) ) \vert {\cal M} \vert^2.
    \label{eq:boltzmannexp}
\end{align}
The ($1+n_B(t,\vec p)$) factor is known as the Bose enhancement factor and is related to potential resonances. We will ignore it in the following calculations. 

Next, we use the identity $\delta(a-b-c)=2b\delta(b^2-(a-c)^2)$ to rewrite the Dirac delta function as
\begin{align}
    \delta ( E^\prime - E - \omega ) = \frac{ E + \omega }{k p} \delta( \cos \theta - \frac{2 E \omega+ \omega^2 - k^2}{2 k p}).
\end{align}
Please take note of that we have adopted the relativistic limit where  $p \gg m$ such that $E \simeq p$.  Putting this into (\ref{eq:boltzmannexp}) along with (\ref{eq: M matrix minimal scenario}), we obtain
\begin{align}
     \frac{{\rm d} \delta f_h(t,k)}{{\rm d} t} \simeq &  \frac{16 G}{4 \omega} \int \frac{\md p \rm dcos \theta }{ 8 E  k p} \delta\left( \cos \theta - \frac{2 E \omega + \omega^2 - k^2}{2 k p} \right) p^4 (1- \cos \theta^2)^2  \frac{1}{e^{(E+\omega)/T}-1} \nonumber \\
     =& \frac{ G}{\omega} \int \frac{\md p}{ 32 E    k^5 } p \left( 4 k^2 p^2 - \left( 2 E \omega+ \omega^2 -k^2\right)^2 \right)^2  \frac{1}{e^{(E+\omega)/T}-1}.
\end{align}
Furthermore, we insert the dispersion relation $\omega = c_T k$ 
 and, therefore, the integration range is restricted to
\begin{align}
    p_{\rm min} = {\rm max}\left( \frac{k (1+c_T)}{2}, m\right).
\end{align}
In later calculations, we will focus on the case where $k > m$, so we can neglect the second limit. We take the upper limit $p_{\rm max} \rightarrow \infty$.
Finally, by solving the integral we find the following analytical form of the Boltzmann equation
\begin{align}
      \frac{{\rm d} \delta f_h(t,k)}{{\rm d} t}  \simeq & \frac{(1-c_T^2)^2 G  T^3 }{32 c_T^3 x^2 } \Big[  (1-c_T^2)^2  x^4  \left( y  + x  -  \log(-1 + e^{x+y}) \right) - 8 c_T^2   y (y  + x ) \nonumber \\
      & \times (2 c_T^2 y^2  + 2 c_T^2 y x  + (-1+c_T^2)  x^2) \log(1 - e^{- x - y})  + 8 c_T^2  
      \Big( (2 y  + x ) \nonumber \\
      & \times (2 c_T y + (-1 + c_T) x )    (2 c_T y+ (1 + c_T) x )  {\rm Plog}(2, e^{-x - y})  \nonumber \\
      & + 2  \Big( (- x^2 + 3 c_T^2 (2 y + x )^2 )  {\rm Plog}(3,e^{-x-y} ) + 12 c_T^2  (2 y + x ) {\rm Plog}(4,e^{-x-y}) \nonumber \\
      & + 24 c_T^2  {\rm Plog}(5,e^{-x-y} ) \Big) \Big) \Big], 
      \label{eq:differentialfh}
\end{align}
where $x= c_T k/T$ and $y=p_{\rm min}/T$, and Plog$(s,z)$ represents the polylogarithm function.

\subsection{Relic Graviton Background in the Minimal Scenario}\label{Sec:Relic Graviton Background}

We are interested in getting the relic graviton background from a distribution of scalar particles after inflation, taking cosmological evolution into account. Assuming a flat background, the differential graviton energy density is given by
\begin{equation}
    d\rho_g(t,k) = 2\omega f_h(t,k) \frac{d^3k}{(2\pi)^3}.
\end{equation}
Generalizing to an expanding universe and integrating over the energy spectrum, this becomes
\begin{align}
    (\partial_t + 4 H) \rho_{g}(t) =& \int \frac{\md^3 k}{(2\pi)^3} 2 \omega  \delta \dot f_h,
    \label{eq:energydensityg}
\end{align}
where $H$ is the Hubble parameter. Following the procedure of \cite{Ghiglieri:2015nfa,Ghiglieri:2022rfp}, we convert the time dependence in (\ref{eq:energydensityg}) to a temperature dependence and redshift all relevant quantities to today. 
Then, we obtain the fractional energy density as
\begin{align}
    h_0^2 \Omega_{g}(k) \simeq & \frac{15 \sqrt{45} }{4 \pi^{11/2}} M_{\rm pl} g_{\star,s}(T_{\rm today})^{1/3} h_0^2 \Omega_\gamma \left( \frac{k }{T_{\rm today}} \right)^3 \nonumber \\
    & \times \int_{T_{\rm min}}^{T_{\rm max}} \frac{{\rm d} T}{T^4} \frac{g_{\star,c}(T)}{g_{\star \rho}(T)^{1/2} g_{\star s}(T)^{4/3}} R\left(T,  \frac{k T}{T_{\rm today}} \left( \frac{g_{\star s}(T)}{g_{\star s}(T_{\rm today})} \right)^{1/3} \right)
\end{align}
where $g_{\star s}$, $g_{\star c}$, and $g_{\star \rho}$ are the effective degrees of freedom for the entropy density, heat capacity, and energy density, respectively, $M_{\rm pl}$ is the Planck mass, $h_0^2 \Omega_\gamma=2.473 \times 10^{-5}$ is the fractional photon energy density today, and $T_{today}=2.725 \rm K$ is the current temperature. We assume that the standard cosmology still applies and $R(t,k) \equiv 2 \omega  \delta \dot f_h(t,k)$. $T_{\rm min}$ is the minimum temperature that can be reached before our assumption that $p \gg m$ no longer holds. When plotting, we set $T_{\rm min}=10^3 m$. $T_{\rm max}$ is the temperature of the distribution of scalar particles at our initial time.

Note that, due to the shift $k \rightarrow k T/T_{\rm today} (g_{\star s}(T)/g_{\star s}(T_{\rm today}))^{1/3}$, the variable $x$ defined below (\ref{eq:differentialfh}) is not time dependent. The same can be said for $y$ as long as $k/a > m $ such that we can define a lower limit on the frequency of the gravitational wave $k_{\rm min} = m a(T_{\rm min})$.
Finally, assuming the effective relativistic degrees of freedom do not change in the temperature range $T_{\rm min} \leq T \leq T_{\rm max}$, the temperature dependence in $R$ factors out as $R \propto T^4$, so the final integrand no longer depends on $T$. 
Furthermore, we approximate $g_{\star s}(T)=g_{\star c}(T) = g_{\star \rho}(T)$ with $T=T_{\rm max}$. Therefore, the fractional energy density is now
\begin{align}
     h_0^2 \Omega_{g}(k) \simeq & \frac{15 \sqrt{45} }{4 \pi^{11/2}} \frac{g_{\star s}(T_{\rm today})^{1/3}}{g_{\star s}(T_{\rm max})^{5/6}}M_{\rm pl} h_0^2 \Omega_{\gamma} \left( \frac{k}{T_{\rm today}} \right)^3 \left( \frac{R(T,  \frac{ k Tg_{\star s}(T_{\rm max})^{1/3}}{T_{\rm today} g_{\star s}(T_{\rm today})^{1/3}})}{T^4} \right) \left( T_{\rm max} - T_{\rm min} \right) \nonumber \\
     \simeq & \frac{15 \sqrt{45} }{4 \pi^{11/2}} \frac{g_{\star s}(T_{\rm today})^{4/3}}{g_{\star s}(T_{\rm max})^{11/6}}  h_0^2 \Omega_{\gamma}  \frac{T_{\rm max}}{M_{\rm pl}} \psi(x_0,y_0) 
\end{align}
with
\begin{align}
    \psi(x_0,y_0) \simeq & \frac{(1-c_T^2)^2 x_0^2 }{16 c_T^6 } \Big[  (1-c_T^2)^2  x_0^4  \left( y_0  + x_0  -  \log(-1 + e^{x_0+y_0}) \right) - 8 c_T^2   y_0 (y_0  + x_0 ) \nonumber \\
      & \times (2 c_T^2 y_0^2  + 2 c_T^2 y_0 x_0  + (-1+c_T^2)  x_0^2) \log(1 - e^{- x_0 - y_0})  + 8 c_T^2  
      \Big( (2 y_0  + x_0 ) \nonumber \\
      & \times (2 c_T y_0 + (-1 + c_T) x_0 )    (2 c_T y_0 + (1 + c_T) x_0 )  {\rm Plog}(2, e^{-x_0 - y_0})  \nonumber \\
      & + 2  \Big( (- x_0^2 + 3 c_T^2 (2 y_0 + x_0 )^2 )  {\rm Plog}(3,e^{-x_0-y_0} ) + 12 c_T^2  (2 y_0 + x_0 ) {\rm Plog}(4,e^{-x_0-y_0}) \nonumber \\
      & + 24 c_T^2  {\rm Plog}(5,e^{-x_0-y_0} ) \Big) \Big) \Big] 
\end{align}
where we used that $T_{\rm max} \gg T_{\rm min}$ and
\begin{align}
    x_0 = \frac{c_T k}{T_{\rm today}} \left(\frac{g_{\star s}(T_{\rm max})}{g_{\star s}}\right)^{1/3}, \qquad y_0  = \frac{(1+c_T) }{2 c_T} x_0
\end{align}
Again, we assume that we are only considering the frequency range $k$ such that $k_{\rm min} \geq m a(T_{\rm min})$.

Firstly, we note that deep in the infrared regime $x_0 \ll 1$, so we get $\Omega_{\rm GW} \propto x_0^2$. To fulfill the BBN bound of $h_0^2\Omega_{\rm g} \lesssim 1.12\times 10^{-6}$, we need 
\begin{align}
    h_0^2\Omega_{g} \simeq 10^{-6}  \frac{T_{\rm max}}{M_{\rm pl}} \psi(x_0) \lesssim 10^{-6}
\end{align}
which requires that $\psi(x_0)  \ll M_{\rm pl}/T_{\rm max}$. For small deviations from GR, i.e. $1-c_T\simeq 0$, this is always fulfilled. 

\begin{figure}[h]
    \centering
    \includegraphics[scale=0.4]{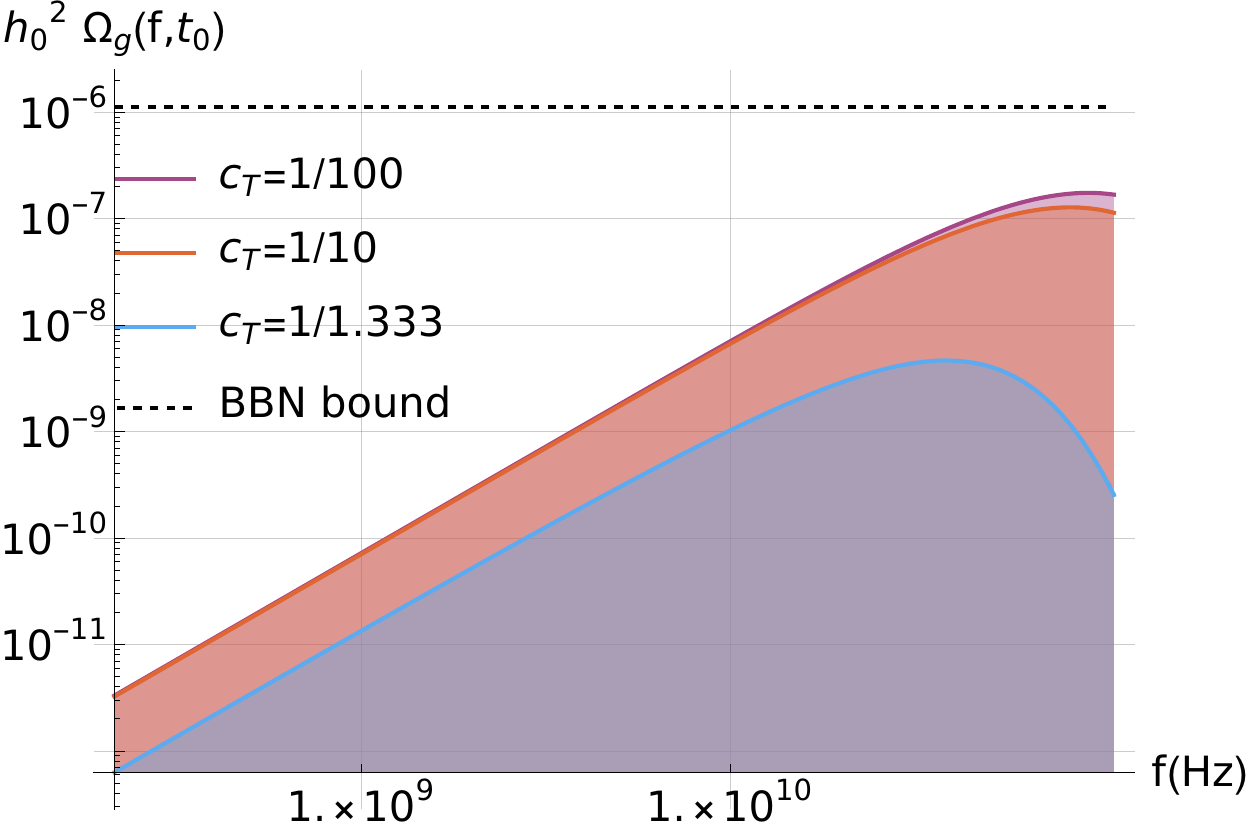}
    \includegraphics[scale=0.4]{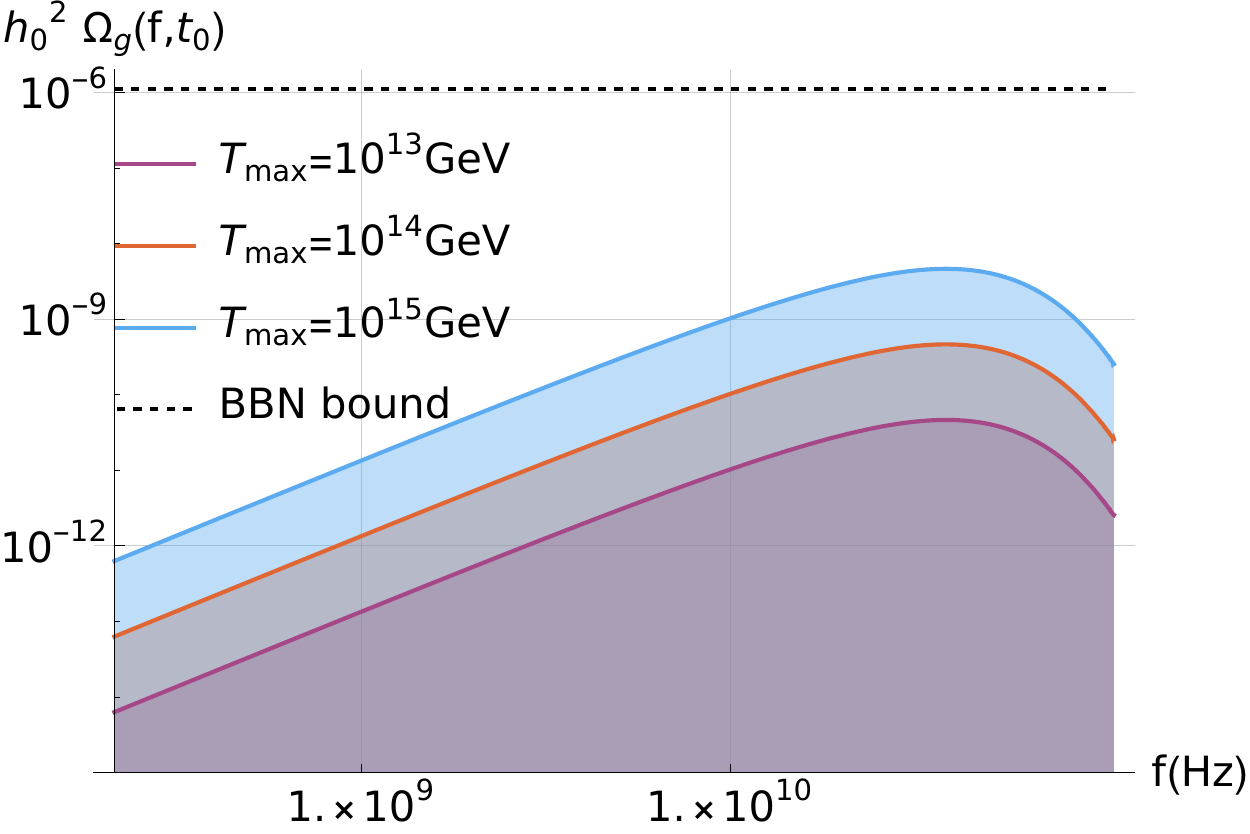}
    \caption{Graviton background for the minimal scenario, i.e. assuming a gravitational wave speed modification only. In these plots, $m=0.1{\rm GeV}$, $g_{\star s}(T_{\rm today})=3.931$ and $g_{\star s}(T_{\rm max})=1$. In the first plot $T_{\rm max}=10^{15}{\rm GeV}$, and in the second 
    $c_T=1/1.333$. The BBN bound is plotted for comparison.}
    \label{fig:minback}
\end{figure}

To see the maximal effect, let us expand around the small $c_T$ limit. 
Defining $\tilde x_0 = x_0/c_T$, we expand $\psi(x_0)$ around $c_T \simeq 0$ leading to
\begin{align}
    \psi(\tilde x_0) \simeq  2 \tilde x_0^2 \left( \tilde x_0^2 {\rm Plog}(3, e^{-\tilde x_0/2}) + 6 \tilde x_0 {\rm Plog}(4, e^{-\tilde x_0/2} )+ 12 {\rm Plog}(5, e^{-\tilde x_0/2}) \right)
\end{align}
which has a maximum around $\psi(\tilde x_0) \simeq 300$ at $\tilde x_0 \simeq 6$. Note, that in the limit $c_T \rightarrow 0$ essentially every particle can contribute to the GW background. However, the slower IR modes are subdominant in comparison to the UV modes and, therefore, the final GW energy spectrum becomes independent of $c_T$ (see Fig. \ref{fig:minback}). The relic graviton background agrees with the BBN bound in most cases. If we  assume that inflation was driven by a single canonical scalar field and the following reheating occurred instantly, we have $ T_{\rm max}/M_{\rm pl} \lesssim 10^{-3}$ obtained from the non-observation of gravitational waves. In light of this order of magnitude estimate, we may naively expect that the relic background may reach the level which is comparable to and thus constrained by the BBN bound in this most optimistic scenario. However, a detailed numerical computation shows that the relic background still agrees with the BBN bound, see Fig.  \ref{fig:minback}. Regardless of the choice of parameters, the background remains below the BBN bound, indicating that no constraints can be imposed on $c_T$ at this point.

So far, we have considered that the universe is dominated by scalar particles in the early Universe. We can generalize the approach to discuss all of the SM particles, similar to \cite{Ghiglieri:2020mhm}, which will all emit Cherenkov radiation if $c_T < 1$. However, we expect that the result will not change significantly. Therefore, it is not possible to constrain $c_T$ by the emission of gravitons at the early Universe due to the Cherenkov radiation of the relativistic particles at this stage.

\section{Horndeski theory}\label{Sec: Quartic Horndeski}
We assume that the modifications to gravity responsible for $c_T \neq 1$ are generated by a non-minimally coupled scalar field, which can be described by Horndeski theory:
\begin{align}
    {\cal L} = & \sqrt{-g} \Big[  G_2 - G_{3}\Box \phi + G_{4} R + G_{4,X} \left( (\Box \phi)^2 - (\nabla_\mu \nabla_\nu \phi)^2 \right) \nonumber \\
    + & G_{5} G^{\mu\nu} \nabla_\mu \nabla_\nu \phi - \frac{1}{6} G_{5,X} \left( (\Box \phi)^3 - 3 \Box \phi (\nabla_\mu \nabla_\nu\phi)^2 + 2 (\nabla_\mu \nabla_\nu \phi)^3 \right) \Big], 
\end{align}
where the free functions $G_i$ depend on $\phi$ and $X= -1/2 \partial_\mu \phi \partial^\mu \phi$. Furthermore, $R$ is the Ricci scalar and $G^{\mu\nu}$ the corresponding Einstein tensor. 

\subsection{Second and cubic order action}
Horndeski theory has been studied in detail in the literature (see \cite{Kobayashi:2019hrl} for a review). In the unitary gauge, the metric and scalar field perturbations are given by $\delta \phi =0$ and
\begin{align}
    \md s^2 = - (1 + \delta N)^2 \md t^2 + a^2 e^{2 \xi } (e^{\gamma})_{ij} ( N^i \md t + \md x^i ) ( N^j  \md t + \md x^j ) 
\end{align}
where $N_i = \partial_i \beta$. The second order action for the scalar and tensor modes is then given by
\begin{align}
    S^{(2)} = \int \md^4x\, \frac{a^3}{8} \left( {\cal G}_T \dot \gamma_{ij}^2 - {\cal F_T} a^{-2} (\partial_k \gamma_{ij})^2 \right) + a^3 \left( {\cal G}_S \dot \zeta^2 - a^{-2} {\cal F}_S (\partial_k \zeta)^2 \right),
\end{align}
and the scalar-scalar-tensor interactions by \cite{Gao:2012ib} 
\begin{align}\label{eq:sshaction}
    {\cal L}_{ssh} &=  \frac{a}{2} \dot \gamma_{ij} \partial_{i}\partial_j\beta \left( \frac{\Gamma \mathcal{G}_t}{\Theta} \dot \zeta - 3 \mathcal{G}_t \zeta \right) - a \gamma_{ij} \left(  \mathcal{F}_t \partial_i\zeta \partial_j\zeta + \frac{2 \mathcal{G}_t ^2}{\Theta} \partial_i\zeta  \partial_j\dot\zeta - \frac{\mathcal{G}_T}{4 a^2} \Delta (\partial_i\beta \partial_j\beta) \right)\nonumber\\&
    +a\mu\left(\frac{{\cal G}_t}{\Theta}\dot \gamma_{ij}\partial_i\dot\zeta\partial_j\zeta-\dot \gamma_{ij}\partial_i\dot\zeta\partial_j\beta-\frac{{\cal G}_t}{a^2\Theta}\gamma_{ij}\Delta(\partial_i\dot\zeta\partial_j\beta)+\frac{1}{2a^2}\dot \gamma_{ij}\Delta(\partial_i\beta\partial_j\beta)\right)\,,
\end{align}
where $\beta$ reads
\begin{align}
\beta=&\frac{1}{a\mathcal{ G}_t}\left(a^3\mathcal{ G}_s\Delta^{-1}\dot\zeta
-\frac{a\mathcal{ G}_t^2}{\Theta}\zeta\right).
\label{eq:shiftsolution}
\end{align}
The explicit form of the theory coefficients are provided in appendix \ref{app:Explicit_Form}.

However, as discussed in the context of induced gravitational waves, the unitary gauge is not suitable for the calculation of the energy spectrum of gravitational waves, as it can lead to gauge artifacts \cite{Hwang:2017oxa,Domenech:2020xin}. Therefore, following \cite{Domenech:2024drm}, we perform a gauge transformation to the Newtonian gauge as
\begin{align}\label{eq:gauge-uniform-newton}
h_{ij}& \rightarrow h_{ij}-a^{-2}\left(\partial_i \beta\partial_j \beta\right)^{TT}\\
\zeta& \rightarrow \zeta +\frac{1}{4}\Delta^{-1}\left(\dot h_{ij}\partial_i\partial_j \beta\right) + \Delta^{-1} \left( h_{ij} \partial_i \partial_j \left( \zeta + \frac{\Theta}{{\cal G}_T} \beta \right) \right)\,,
\end{align}
where we only wrote down the terms relevant for the cubic scalar-scalar-tensor interaction. Furthermore, we have the linear relation $\beta = \delta \phi/\dot \phi \equiv \pi$ between the shift vector in the unitary gauge and the scalar field in the Newtonian gauge. 

In the following, we are interested in the interactions between the scalar field and the tensor modes deep inside the horizon. Therefore, for our purpose it is sufficient to focus only on the leading terms for $k \gg a H$.

First of all, we can express the scalar second order action in terms of $\pi$ as
\begin{align}
    S_s^{(2)} \simeq \int \md^4x\,a^3 \frac{\Theta^2 {\cal G}_s}{{\cal G}_T^2} \left(  \dot \pi^2 - a^{-2} c_s^2 (\partial_k \pi)^2 \right) \simeq \int \md^4x\, \frac{a^3}{2} \left( (\dot \pi^{(c)})^2 - a^{-2} c_s^2 (\partial_k \pi^{(c)})^2 \right), 
\end{align}
where in the second step we have introduced the canonical normalized variables 
\begin{align}
    \pi^{(c)} = \sqrt{\frac{2 {\cal G}_s \Theta^2}{ {\cal G}_T^2}}\, \pi.
\end{align}
Additionally, we used the relation 
\begin{align}
    \zeta \simeq - \frac{\Theta}{{\cal G}_T} \pi - \frac{{\cal G}_s \Theta^2}{{\cal G}_T^3} \frac{a^2}{\Delta} \dot \pi
\end{align}
(see appendix \ref{app:Integrating_pi} for more details about obtaining the second order action in terms of $\pi$).
As we are deep inside the horizon, we will set $a=1$ in the following. The tensor modes are gauge invariant up to linear order and, therefore, the second order action does not change and the canonical normalized tensor modes are given by
\begin{align}
    \gamma_{ij}^{(c)} = \sqrt{\frac{{\cal G}_T}{4}} \gamma_{ij}.
\end{align}

Finally, the leading order terms for the cubic interaction in the Newtonian gauge can be expressed as
\begin{align}
    {\cal L}_{ssh} \simeq & \frac{ {\cal G}_T^2}{2 \Theta} \left( 1 - \frac{\Gamma}{{\cal G}_T} \right) \dot \gamma_{ij} \partial_i \pi \partial_j \dot \zeta + \frac{{\cal G}_T}{4} (1-c_T^2) \Delta ( \partial_i \pi \partial_j \pi ) + \mu \left( - \gamma_{ij} \Delta ( \partial_i \dot \pi \partial_j \pi) - \frac{{\cal G}_T }{\Theta} \gamma_{ij} \Delta( \partial_i \dot \zeta \partial_j \pi) \right) \nonumber \\
    \simeq & - \frac{{\cal G}_T }{4} \left(1 - \frac{\Gamma}{{\cal G}_T} \right) \gamma_{ij} \frac{\md^2}{\md t^2} \left( \partial_i \pi \partial_j \pi \right) + \frac{{\cal G}_T}{4} (1-c_T^2) \gamma_{ij} \Delta (\partial_i \pi \partial_j \pi) + \frac{\mu {\cal G}_T}{\Theta} \frac{\md }{\md t} \left( \frac{\Theta}{{\cal G}_T} \right) \gamma_{ij} \Delta (\partial_i \pi \partial_j \pi ) \nonumber \\
    & + \frac{{\cal G}_s \Theta}{{\cal G}_T^2 } \gamma_{ij} \Delta \left( \frac{\partial_i \ddot \pi}{\Delta} \partial_j \pi \right).
\end{align}
We can further simplify this result by using the linear equations of motion and dispersion relation in the $k \gg  H$ limit, i.e. $\ddot \pi \simeq c_s^2 \Delta \pi$ and $\ddot \gamma_{ij} \simeq c_T^2 \Delta \gamma_{ij}$. In this case it is straightforward to see that we can simplify all the terms into one single cubic interaction:
\begin{align}
    {\cal L}_{ssh} \simeq \Big[ -\frac{{\cal G}_T c_T^2}{4} \left(1 - \frac{\Gamma}{{\cal G}_T} \right) + \frac{{\cal G}_T}{4} (1-c_T^2) + \frac{\mu {\cal G}_T}{\Theta} \frac{\md}{\md t} \left( \frac{\Theta}{{\cal G}_T} \right) + \frac{{\cal G}_s c_s^2 \Theta}{{\cal G}_T^2} \Big] \frac{{\cal G}_T^{3/2}}{ {\cal G}_s \Theta^2 c_T^2} \ddot \gamma_{ij}^{(c)}  \partial_i \pi^{(c)} \partial_j \pi^{(c)}.
\end{align}
Note that the interaction has the same form as the one considered in \cite{Creminelli:2018xsv}, where the authors discussed the decay of gravitons into scalar particles in the case of Beyond Horndeski with $c_T=1$.
For $k \gg H$, there is only one independent term with four derivatives acting on the perturbations. The other terms can be brought into the same form by using the linear equations of motion. 

Furthermore, we also note that the leading interaction for the non-minimally coupled Horndeski scalar field has two more time derivatives acting on the tensor modes than the standard Cherenkov radiation from minimally coupled scalar fields, which scales as ${\cal L}^{\rm min}_{ssh} \sim \gamma_{ij} \partial_i \pi \partial_j \pi$. Therefore, as discussed in Appendix \ref{app:General_cs}, the diagram has non-vanishing contributions for $c_T =1$ but $c_s \neq 1$.

The corresponding interaction amplitude is given by
\begin{align}
    i {\cal M} = \frac{2 i}{\Lambda_\star^3} k
    ^2 p_{m} p^\prime_{n} \epsilon^{\sigma}_{mn}(\vec k),
\end{align}
where ${\cal M}$ stands for the Feynman diagram where an incoming scalar particle with momentum $p^\prime$ decays into an outgoing scalar particle with momentum $p$ and a gravitational wave $\gamma_{ij}^{(c)}$, and 
\begin{align}\label{eq:lambdastar}
    \Lambda_\star^3 \equiv \frac{ {\cal G}_s \Theta^2}{{\cal G}_T^{3/2}} \Big[ -\frac{{\cal G}_T c_T^2}{4} \left(1 - \frac{\Gamma}{{\cal G}_T} \right) + \frac{{\cal G}_T}{4} (1-c_T^2) + \frac{\mu {\cal G}_T}{\Theta} \frac{\md}{\md t} \left( \frac{\Theta}{{\cal G}_T} \right) + \frac{{\cal G}_s c_s^2 \Theta}{{\cal G}_T^2} \Big]^{-1} .
\end{align}

Similarly to the approach carried out in Section \ref{Sec:Gravitational Cherenkov Interaction}, we sum over all the polarizations, which leads to 

\begin{align}\label{eq: A squared}
    \langle \vert {\cal M} \vert^2 \rangle = \frac{1}{2} \sum_{\sigma} \vert  i {\cal M} \vert^2 = \frac{2}{\Lambda_\star^6}  k^4 \left( p^2 - \frac{(\vec p \cdot \vec k)^2}{k^2} \right)^2.
\end{align}

\subsection{Relic Graviton Background in Horndeski Theory}

As shown in Section \ref{Sec:Minimal Scenario}, the minimally coupled thermalized particles do not violate the BBN bound and, therefore, do not impose constraints on the gravitational wave speed in the early Universe. For this reason, in the present section we consider the case of a non-minimally coupled thermalized scalar field, which is responsible for the modifications of gravity and which could be, for instance, a remnant of the inflaton during reheating. 

From equations \eqref{eq: M matrix minimal scenario} and \eqref{eq: A squared} we find that the interaction between the graviton and the scalar field in Horndeski is rescaled by 
\begin{align}
    \frac{\langle \vert {\cal M} \vert^2 \rangle_{\rm H} }{\langle \vert {\cal M} \vert^2 \rangle_{\rm min} } \simeq \frac{1}{8\pi G}\frac{k^4 }{\Lambda_\star^6} 
\end{align}
when compared to the minimal scenario. Given that this factor is $k$ dependent, the integrand is modified and acquires a $T^4$ dependence.

Then, computing the ratio between the minimal relic background $h_0^2\Omega_{\rm g, min}$ and the Horndeski relic background $h_0^2\Omega_{\rm g, H}$, we obtain 
\begin{align}\label{eq:qhback}
    \frac{\Omega_{\rm g, H}}{\Omega_{\rm g, min}} \simeq \frac{1}{5 } \frac{T_{\rm max}^4 M_{\rm pl}^2} {8\pi \Lambda_\star^6 }  \tilde x_0^4, 
\end{align}
where we recall that we have neglected any time dependence of the modified gravity parameters and assumed $c_s =1$ for simplicity. In Appendix \ref{app:General_cs}, we show the full expression for a the more general case of $c_s \neq 1$. Note that the spectrum vanishes for $c_s = c_T=1$, as expected due to energy and momentum conservation. However, we still get a non-vanishing contribution for $c_T =1 $ but $c_s < 1$ due to the higher derivative interactions. It is similar to \cite{DeFelice:2015moy}, where the authors considered the same vertex for the decay of a graviton into two scalar particles for $c_s < c_T=1$ in Beyond Horndeski theory.

\begin{figure}[h]
    \centering
    \includegraphics[scale=0.4]{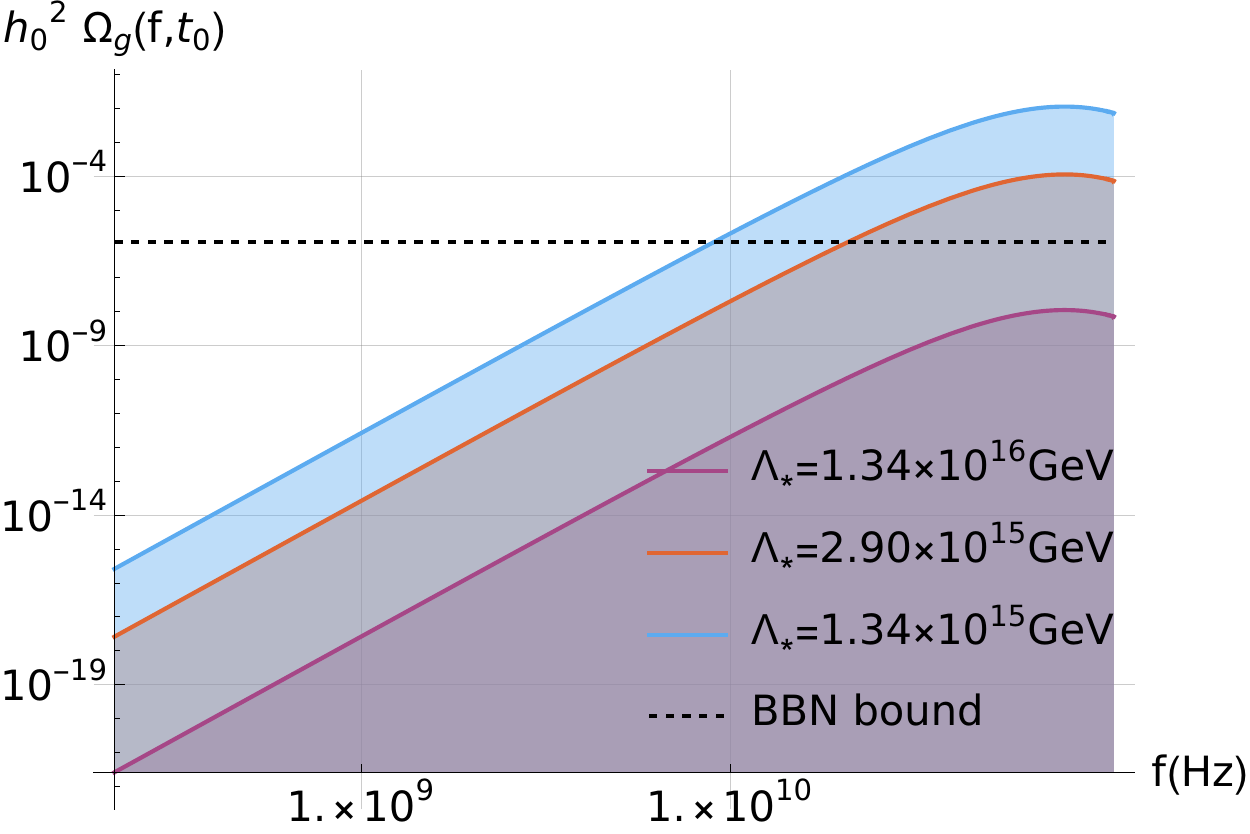}
    \includegraphics[scale=0.4]{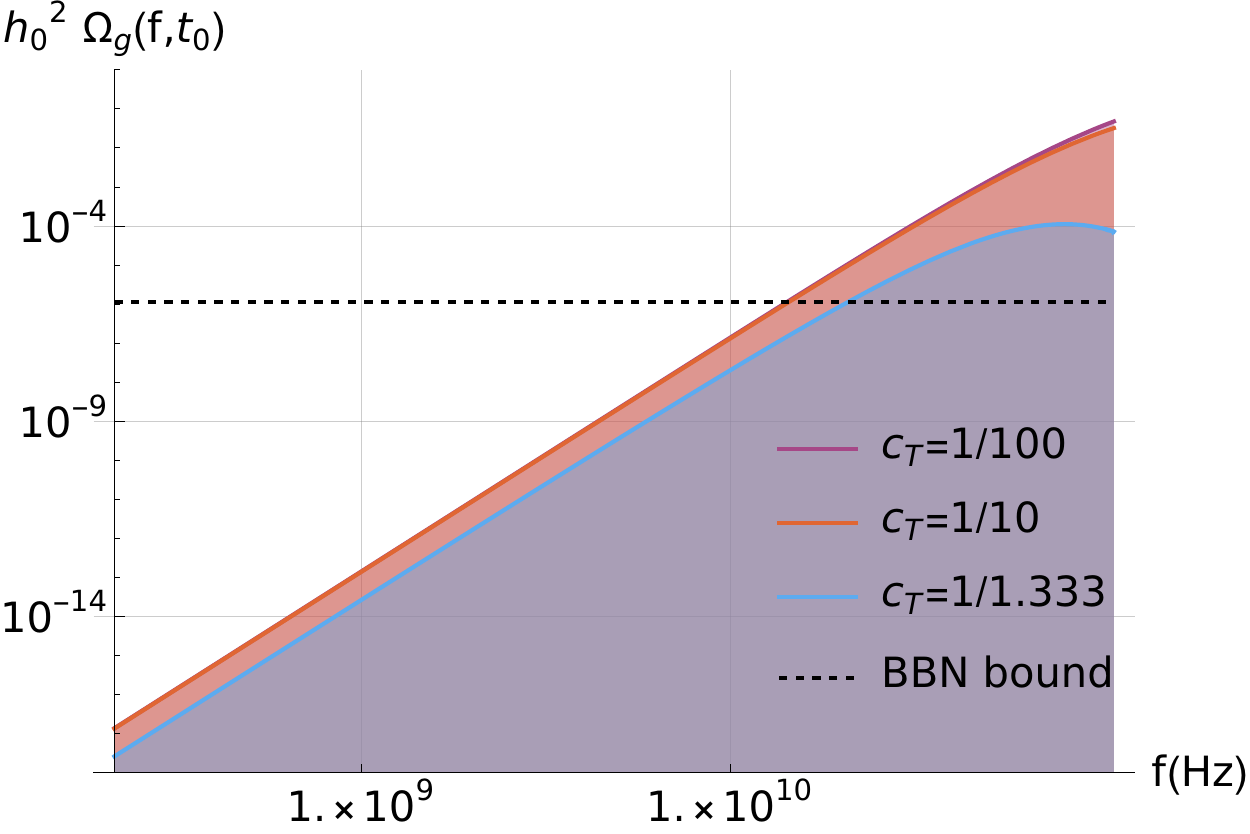}
    
    \caption{Graviton background in a general Horndeski theory with $g_{\star s}(T_{\rm today})=3.931$, $g_{\star s}(T_{\rm max})=1$, $T_{\rm max}=10^{15}{\rm GeV}$, $m=0.1{\rm GeV}$. In the first plot we fix $c_T=1/1.333$, while in the second  $\Lambda_*=2.90\times10^{15}{\rm GeV}$.}
    \label{fig:qH}
\end{figure}

\begin{figure}[h]
    \centering
\includegraphics[scale=0.4]{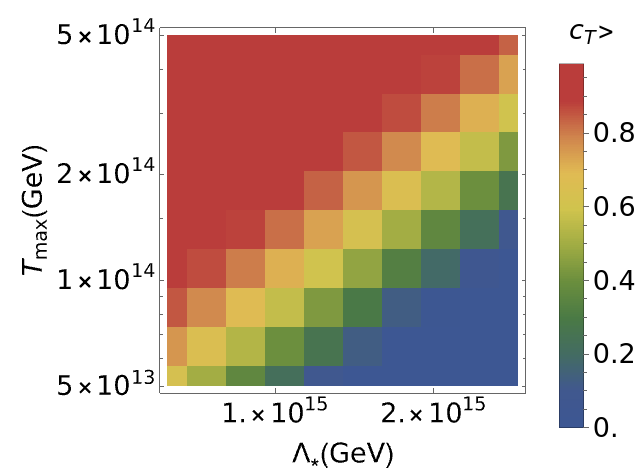}
\includegraphics[scale=0.4]{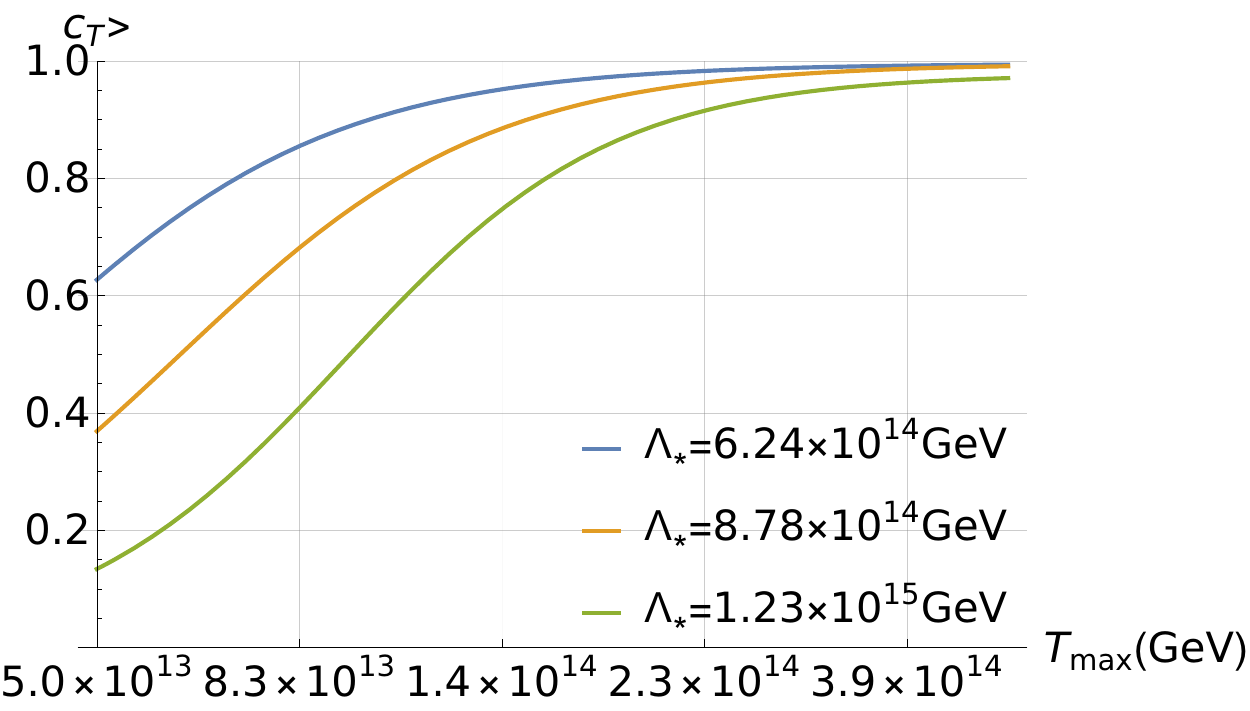}
    \caption{Lower bound on the gravitational wave speed $c_T$ within Horndeski theories as a function of $\Lambda_\star$ and $T_{\rm max}$. In these plots $g_{\star s}(T_{\rm today})=3.931$, $g_{\star s}(T_{\rm max})=1$, and $m=0.1{\rm GeV}$. The bound is set by requiring that the background contribution to the number of effective relativistic species $N_{\rm eff}$ is allowed by BBN. We consider $10$ values of $\Lambda_\star$ and $10$ values of $T_{\rm max}$ in the first plot, which are then fitted for three values of $\Lambda_\star$ and depicted in the second plot.}
\label{fig:ctconstraintsH}
\end{figure}

Figure \ref{fig:qH} shows the relic background in Horndeski theories for different values of $\Lambda_\star$, as well as the BBN bound for comparison. We see that for a region of the parameter space, the BBN bound is violated. Figure \ref{fig:ctconstraintsH} shows the resulting constraints on $c_T$ by requiring that the relic background contribution to the number of relativistic species $N_{\rm eff}$ is in agreement with BBN. Depending on the details of the model, $\Lambda_\star$ is in general not independent from $c_T$, but it is assumed here for simplicity. 

\subsection{Relic Graviton Background in Galileon Theory}

As a Horndeski example let us consider the weakly broken Galilean symmetry \cite{Pirtskhalava:2015nla}. Considering only the leading order terms for the free functions we have
\begin{align}
    {\cal L} =& \sqrt{-g}\Big[ X - V(\phi)  - c_3 X \frac{\Box \phi}{\Lambda_3^3} + \frac{1}{2} M_{\rm pl}^2 R + \frac{c_4 X^2}{\Lambda_3^6}  R + \frac{2 c_4 X }{\Lambda_3^6}  \left( (\Box \phi)^2 - (\nabla_\mu \nabla_\nu \phi)^2 \right) \nonumber \\  & +  c_5 \frac{X^2}{\Lambda_3^9} G^{\mu\nu} \nabla_\mu\nabla_\nu \phi - \frac{c_5}{3} \frac{X}{\Lambda_3^9} \left(  (\Box \phi)^3 - 3 \Box \phi (\nabla_\mu \nabla_\nu\phi)^2 + 2 (\nabla_\mu \nabla_\nu \phi)^3\right) \Big],
\end{align}
where the potential $V(\phi)$ is naturally suppressed as it breaks the Galileon symmetry, and $c_3 \sim c_4 \sim c_5 \sim {\cal O}(1)$. For simplicity, we consider $c_5 = 0$ as the overall behavior does not change by including the quintic contributions. 

Using this ansatz the free functions simplify to 
\begin{align}\label{eq:cTandLambda}
   & {\cal G}_T = M_{\rm pl}^2 - 6 c_4 \frac{X^2}{\Lambda_3^6}, \qquad \Theta = - c_3 \dot \phi \frac{ X}{\Lambda_3^3} +  H M_{\rm pl}^2 -  \frac{30 c_4 X^2 H }{\Lambda_3^6}, \qquad  \Gamma = M_{\rm pl}^2 - \frac{30 c_4 X^2}{\Lambda_3^6}, \nonumber \\
    &    c_T^2 = \frac{1 +  \frac{2 c_4 X^2}{\Lambda_3^6 M_{\rm pl}^2} }{1 -  \frac{6 c_4 X^2}{\Lambda_3^6 M_{\rm pl}^2} } \simeq 1 + 8 c_4 \frac{X^2}{\Lambda_2^8},
\end{align}
where the expansion parameter for the EFT is $X/\Lambda_2^4$, while $M_{\rm pl}^2 = \Lambda_2^8/\Lambda_3^6$. 

Additionally, at next-to-leading order the scalar sound speed is given by
\begin{align}
    c_s^2 \simeq \frac{ M_p^2 \epsilon H^2}{X} \left( 1 + \frac{18 c_3 H \dot \phi}{\Lambda_3^3 } + \frac{c_3 \dot \phi^3 }{2 \epsilon H \Lambda_3^3 M_p^2 }  + \frac{3 c_3 \dot \phi^2 \ddot \phi }{2 \epsilon H^2 M_p^2 \Lambda_3^3 } \right),
\end{align}
i.e. one can tune it to $c_s=1$ independently from $c_4$.

The prefactor of the modified energy spectrum then reads
\begin{align}
     - \left(1 - \frac{\Gamma}{{\cal G}_T} \right) c_T^2 + (1-c_T^2) =  8 c_4 \frac{X^2}{\Lambda_2^8} \frac{-32 \frac{X^2}{\Lambda_2^8}}{(1-6 c_4 \frac{X^2}{\Lambda_2^8})^3} \simeq -32 c_4 \frac{X^2}{\Lambda_2^8}.
\end{align}

Finally, by using ${\cal G}_S \simeq X/H^2 + {\cal O}(X/\Lambda_2^4) $, we obtain
\begin{align}
    \Lambda_\star^3 \simeq 2 \frac{X}{\Lambda_2^4} \Lambda_3^3.
\end{align}

The relic graviton background can be expressed as
\begin{align}\label{eq:Gback}
    \frac{\Omega_{\rm g,G}}{\Omega_{\rm g,min}} \simeq \frac{32 c_4^2}{5\pi } \frac{T_{\rm max}^4 M_{\rm pl}^4}{\Lambda_2^8} \frac{X^2}{\Lambda_2^8} \tilde x_0^4,
\end{align}
where $\Omega_{\rm g,G}$ stands for the background in Galileon theory. 

\begin{figure}[h]
    \centering
    \includegraphics[scale=0.4]{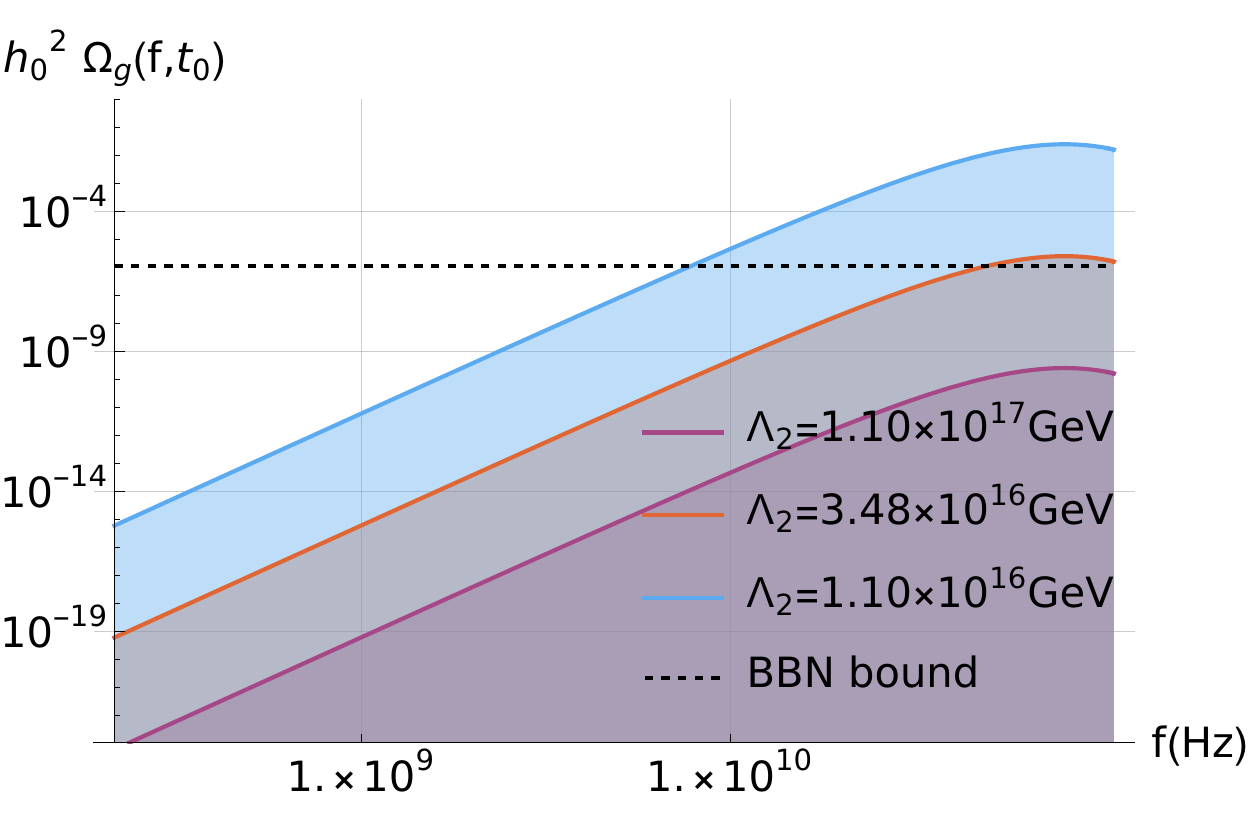}
    \includegraphics[scale=0.4]{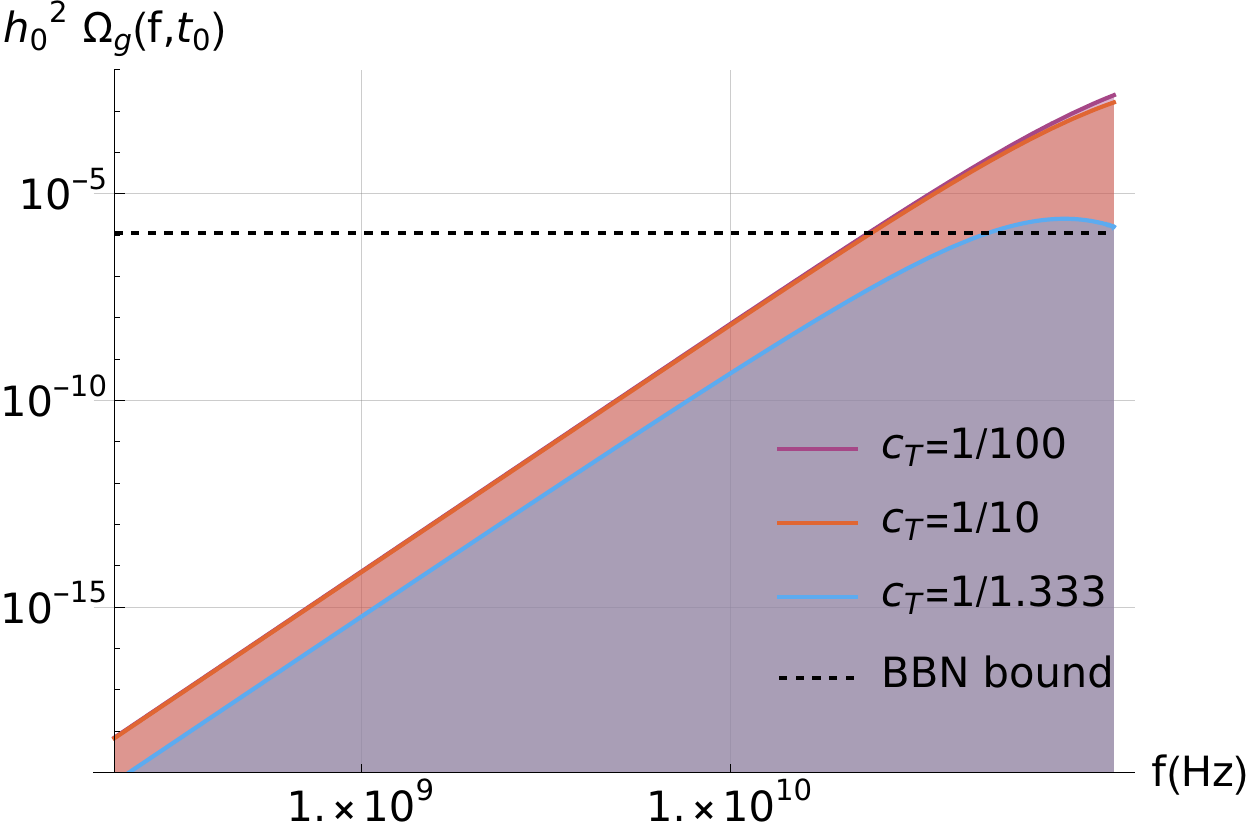}
    \caption{Graviton background in Galileon theory with $T_{\rm max}=10^{15}{\rm GeV}$, $g_{\star s}(T_{\rm today})=3.931$, $g_{\star s}(T_{\rm max})=1$, $m=0.1{\rm GeV}$ and $c_4=-1$. In the first plot we fix $c_T=1/1.333$, while in the second  $\Lambda_2=3.48\times10^{16}{\rm GeV}$.}
    \label{fig:backgroundgalileon}
\end{figure}

\begin{figure}[h]
    \centering
    \includegraphics[scale=0.4]{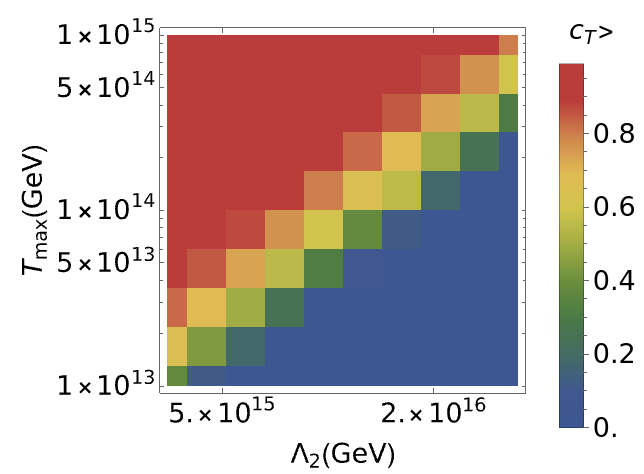}
    \includegraphics[scale=0.4]{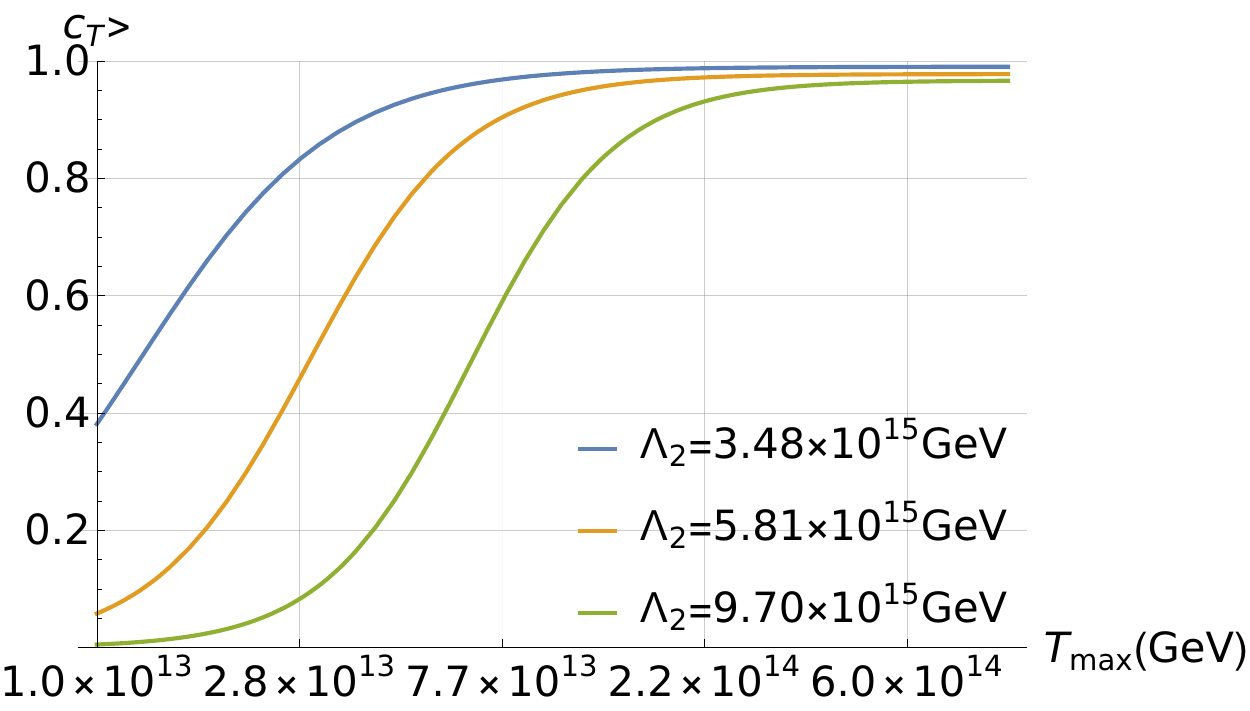}
    \caption{Lower bound on the gravitational wave speed $c_T$ within Galileon thoery as a function of $\Lambda_2$ and $T_{\rm max}$. In these plots $g_{\star s}(T_{\rm today})=3.931$, $g_{\star s}(T_{\rm max})=1$, $m=0.1{\rm GeV}$, and $c_4=-1$. The bound is set by requiring that the background contribution to the number of effective relativistic species $N_{\rm eff}$ is allowed by BBN. We consider $10$ values of $\Lambda_2$ versus $10$ values of $T_{\rm max}$ in the first plot, which are then fitted for three values of $\Lambda_2$ and depicted in the second plot.}
    \label{fig:ctconstraintsgalileon}
\end{figure}

Figure \ref{fig:backgroundgalileon} shows the relic graviton background in the specific case of Galileon theory for different values of $\Lambda_2$ and $c_T$, where we have assumed $c_s=1$. In agreement with the results of the general case of Horndeski theories, we obtain that the BBN bound is violated for a region of the parameter space. Figure \ref{fig:ctconstraintsgalileon} presents the lower bound on $c_T$ that is set by requiring that the background remains below the BBN bound. 

In this case, we have not taken strong coupling into account. However, as long as $k,\, T_{\rm max} \ll \Lambda_3$, this is not an issue.

\section{Discussion and Conclusions}\label{Sec:Conclusion}

In this paper, we have investigated the gravitational Cherenkov radiation originating from a thermal scalar field traveling faster than gravitational waves in the early Universe and obtained the corresponding relic graviton background. 

We started by considering a minimal scenario where only $c_T$ is modified, without specifying the theory that leads to the subluminal behavior. In this case, the graviton background sourced by a minimally coupled scalar field remains below the BBN bound, leading to no constraints on $c_T$. 

We proceeded to include the interaction terms that arise in Horndeski theory and obtained that the gravitational constant is replaced by a factor that depends on the modified gravity theory details, in particular the scale $\Lambda_\star^3$ defined in equation \eqref{eq:lambdastar}, and the frequency of the gravitational wave. This replacement can lead to a significant increase of the background for a region of the parameter space, violating the BBN bound and, therefore, leading to constraints on $c_T$, as shown in Figure \ref{fig:ctconstraintsH}. Furthermore, we exemplify this scenario by considering a specific model, namely the Galileon theory, which leads to the constraints on $c_T$ presented in Figure \ref{fig:ctconstraintsgalileon}.

Note that it is possible to tune the free functions of (beyond) Horndeski theory such that the higher order cubic interactions are still present if $c_T=1$ but $c_s < 1$, similar to \cite{DeFelice:2015moy, Domenech:2024drm}, which will still lead to a sizeable relic graviton background, as discussed in Appendix \ref{app:General_cs}. 
However, due to the presence of the higher derivative terms, the emission of gravitons differs significantly from the standard Cherenkov radiation of a minimally coupled scalar field, which vanishes for $c_s < c_T=1$. 

Additionally, we emphasize that the graviton production discussed here occurs after inflation, i.e. an eventual detection of the relic graviton background could be due to the gravitational Cherenkov production, even if the primordial graviton blackbody has been previously diluted by inflation \cite{Vagnozzi:2022qmc}. Therefore, in order to rule out an inflationary phase due to the observation of a relic graviton background, one would have to rely on the spectral differences between the gravitational Cherenkov and the graviton blackbody radiation. 

Finally, it is worth mentioning that we allow $R(t,k)>H$, i.e. the interaction rate to be larger than the expansion rate, which leads to the thermalization of the produced gravitons. We only limit the production rate so as to certify that $h_0^2\Omega_g\ll 1$. The graviton freeze-out then takes place as soon as the system reaches the minimum tempeature $T_{\rm min}$ and the gravitational Cherenkov interaction stops, which is set by the mass $m$ of the scalar particle. 

Our results represent an important step in the direction of testing gravitational wave speed in the early Universe, establishing significant constraints to $c_T$ in the context of Horndeski theories. 

Notably, we found that the relic graviton background can exceed the BBN bound by quite a few orders of magnitude, a result rarely achieved by other graviton production processes \cite{Giovannini:2019oii, Ghiglieri:2022rfp,Giovannini:2014jca,Vagnozzi:2022qmc,Ringwald:2020ist}.

In future works, it may be interesting to generalize our approach to consider any non-minimally coupled degree of freedom which can generate Cherenkov radiation.

\section*{Acknowledgements}

We would like to thank Federico Urban, Jan Tr\"ankle and Guillem Dom\`enech for helpful discussions. P.C.M.D. acknowledges support from MEYS through the INTER-EXCELLENCE II, INTER-COST grant LUC23115. A.G. is supported by the DFG under the Emmy-Noether program, project number 496592360, and by the JSPS KAKENHI grant No. JP24K00624. C.L. and R.T. are supported by the grant No. 2021/42/E/ST9/00260 from the National Science Centre, Poland.

\appendix

\section{Explicit form of the coefficients}
\label{app:Explicit_Form}

In the following, we list all the coefficients appearing in the second and cubic order action as derived by Ref.~\cite{Gao:2012ib}. First, the coefficients for the second order action for tensor modes and scalar modes are given by
\begin{align}
\mathcal{ F}_t=&2\left(G_4
-X\left( \ddot\phi G_{5X}+G_{5\phi}\right)\right),
\\
\mathcal{ G}_t=&2\left(G_4-2 XG_{4X}
-X\left(H\dot\phi G_{5X} -G_{5\phi}\right)\right), \\
\mathcal{ F}_s=&\frac{1}{a}\frac{d}{dt}\left(\frac{a}{\Theta}\mathcal{ G}_t^2\right)
-\mathcal{ F}_t,
\\
\mathcal{ G}_s=&\frac{\Sigma }{\Theta^2}\mathcal{ G}_t^2+3\mathcal{ G}_t,
\end{align}
where $\Sigma$ and $\Theta$ are defined by
\begin{align}
\Sigma=&XG_{2,X}+2X^2G_{2,XX}+12H\dot\phi XG_{3X}
+6H\dot\phi X^2G_{3XX}
-2XG_{3\phi}-2X^2G_{3\phi X}-6H^2G_4\nonumber\\&
+6\left[H^2\left(7XG_{4X}+16X^2G_{4XX}+4X^3G_{4XXX}\right)
-H\dot\phi\left(G_{4\phi}+5XG_{4\phi X}+2X^2G_{4\phi XX}\right)
\right]\nonumber\\&
+30H^3\dot\phi XG_{5X}+26H^3\dot\phi X^2G_{5XX}
+4H^3\dot\phi X^3G_{5XXX} \nonumber\\&
-6H^2X\left(6G_{5\phi}
+9XG_{5\phi X}+2 X^2G_{5\phi XX}\right),
\\
\Theta=&-\dot\phi XG_{3X}+
2HG_4-8HXG_{4X}
-8HX^2G_{4XX}+\dot\phi G_{4\phi}+2X\dot\phi G_{4\phi X}\nonumber\\&
-H^2\dot\phi\left(5XG_{5X}+2X^2G_{5XX}\right)
+2HX\left(3G_{5\phi}+2XG_{5\phi X}\right).
\end{align}

In the third order action for scalar-scalar-tensor interactions, we have that
\begin{align}
\mu=& \dot\phi XG_{5X}\,, \\
\Gamma=& 2G_4-8XG_{4X}-8X^2G_{4XX}
-2H\dot\phi\left(5XG_{5X}+2X^2G_{5XX}\right)
+2X\left(3G_{5\phi}+2XG_{5\phi X}\right).
\end{align}

\section{Integrating in $\pi$}
\label{app:Integrating_pi}

To obtain the second order Lagrangian in terms of $\pi$ we can follow  \cite{DeFelice:2015moy}, including time dependent coefficients. 
The original Lagrangian in Fourier space is given by 
\begin{align}
    \mathcal{L}^{(2)} = a^3 \Big[ {\cal G}_s \dot \zeta^2 - {\cal F}_s \frac{k^2}{a^2} \zeta^2 \Big] \equiv \frac{1}{2} \Big[ A \dot \zeta^2 - B \zeta^2 \Big],
\end{align}
while the relation between $\pi$ and $\zeta$ reads
\begin{align}
   \pi = - \frac{1}{a\mathcal{ G}_t}\left(a^3\mathcal{ G}_s k^{-2} \dot\zeta
+ \frac{a\mathcal{ G}_t^2}{\Theta}\zeta\right) \equiv C \dot \zeta + D \zeta,
\end{align}
where we have introduced $A$, $B$, $C$ and $D$ to shorten the notation. 
To integrate in $\pi$ we can note that the Lagrangian 
\begin{align}
    \tilde {\cal L} =& \frac{A}{2 C^2} \left(2 \pi (C \dot \zeta + D \zeta) - \pi^2 \right) - \frac{A D^2 + B C^2 - C^2 \partial_t{(A D/ C)} }{2 C^2} \zeta^2 \nonumber \\
    =& \frac{A}{2 C^2} \left( 2 C \zeta \dot \pi + \left( 2 D -  \frac{2C^2}{A} \frac{\md}{\md t}\left(\frac{A}{C} \right)  \right) \zeta \pi - \pi^2  \right) - \frac{A D^2 + B C^2 - C^2 \partial_t{(A D/ C)} }{2 C^2} \zeta^2
\end{align}
is equivalent to the original Lagrangian ${\cal L}$ by integrating out $\pi$ up to total derivatives. 
Instead, we can integrate out $\zeta$ via 
\begin{align}
    \zeta = \frac{ ( A D  - C \dot A  + A \dot C) \pi - A C \dot \pi }{B C^2 - C D \dot A + A ( D ( D + \dot C) - C \dot D)},
\end{align}
arriving finally at a Lagrangian purely in terms of $\tilde {\cal L}(\pi, \dot \pi)$. The full expression is quite lengthy. Therefore, we will not explicitly write it down as it is not relevant for our purpose but instead directly expand it for $k \gg a H$, noting that $B \propto k^2$ and $C \propto k^{-2}$, which leads to
\begin{align}
    \tilde {\cal L} \simeq a^3 \frac{\Theta^2 {\cal G}_s}{{\cal G}_T^2} \left(  \dot \pi^2 - a^{-2} c_s^2 (\partial_k \pi)^2 \right). 
\end{align}
Furthermore, the relation between $\zeta$ and $\pi$ in the limit $k \gg H$ can be expressed as
\begin{align}
   \zeta \simeq  - \frac{\Theta}{{\cal G}_T} \pi - \frac{{\cal G}_s \Theta^2}{{\cal G}_T^3} \frac{a^2}{\Delta} \dot \pi.
\end{align}

\section{General $c_s \neq 1$}
\label{app:General_cs}
In the case of $c_s \neq 1$, we obtain $E \simeq c_s p$ in the ultra-relativistic limit. Using the adapted dispersion relation we obtain the Boltzmann function
\begin{align}
    \frac{{\rm d} \delta f_h(t,k)}{{\rm d} t}  \simeq & \frac{c_T   T^7 }{256 c_s^6  \pi \Lambda_\star^6 } x^2 \Big[ \frac{c_s^4}{c_T^4}  (1-c_T^2)^4  x^4  \left( y  + x  -  \log(-1 + e^{x+y}) \right) - \frac{8}{c_T^2}  y (-y+c_s^2 (c_T^2 y + (-1+c_T^2) x ) ) \nonumber \\
    &\times (-2 c_T^2 y^2 + c_s^2 (x^2 (-1+c_T^2)^2 + 2 c_T^2 (-1+c_T^2) x y + 2 c_T^4 y^2  ) ) \log (1 - \cosh (x+y) + \sin(x+y) )  \nonumber \\
      &  + \frac{8}{c_T^2} (c_s (-1+c_T^2) x + 2 c_T (-1 + c_s c_T) y ) (c_s (-1+c_T^2) x + 2 c_T (1+c_s c_T) y ) \nonumber \\
      & \times  (-2 y + c_s^2 ( (-1+c_T^2 ) x + 2 c_T^2 y )   
       {\rm Plog}(2, e^{-x - y}) + 16 c_s^2 \Big( \frac{(-1+c_T^2)^2 (-1+3 c_s^2 c_T^2)x^2}{c_T^2} \nonumber \\
       & + 12 (-1+c_T^2) (-1+c_s^2 c_T^2) x y + \frac{12 (-1+ c_s^2 c_T^2)^2 y^2 }{c_s^2}  \Big) {\rm Plog}(3,e^{-x-y})  + 192 (-1 + c_s^2 c_T^2) \nonumber \\
      & \times  (-2 y + c_s^2 ((-1+c_T^2) x + 2 c_T^2 y))  {\rm Plog}(4,e^{-x-y}) + 384 (-1 + c_s^2 c_T^2)^2 {\rm Plog}(5,e^{-x-y})  \Big],
\end{align}
where $y= c_s p_{\rm min}/T$ and for $c_s \leq 1$
\begin{align}
    p_{\rm min} = {\rm max} \left( \frac{(1-c_T^2) k}{2 ( 1 +c_s c_T) } , m \right).
\end{align}
Substituting it into the expression for the gravitational wave energy spectrum yields
\begin{align}
    \Omega_{\rm g, H} \simeq \frac{3 \sqrt{15} }{4 \pi^{11/2}} \frac{g_{\star s}(T_{\rm today})^{4/3}}{g_{\star s}(T_{\rm max})^{11/6}} h_0^2 \Omega_\gamma  \frac{T_{\rm max}^5 M_{\rm pl} }{8 \pi \Lambda_\star^6 } \tilde x_0^4 \tilde \psi(x_0,y_0),
\end{align}
where
\begin{align}
    \tilde \psi(x_0,y_0) \simeq &  \frac{1   }{128 c_s^6 c_T^2 \pi  } x_0^6 \Big[ \frac{c_s^4}{c_T^4}  (1-c_T^2)^4  x_0^4  \left( y_0  + x_0  -  \log(-1 + e^{x_0+y_0}) \right) - \frac{8}{c_T^2}  y_0 (-y_0+c_s^2 (c_T^2 y_0 + (-1+c_T^2) x_0 ) ) \nonumber \\
    &\times (-2 c_T^2 y_0^2 + c_s^2 (x_0^2 (-1+c_T^2)^2 + 2 c_T^2 (-1+c_T^2) x_0 y_0 + 2 c_T^4 y_0^2  ) ) \log (1 - \cosh (x_0+y_0) + \sin(x_0+y_0) )  \nonumber \\
      &  + \frac{8}{c_T^2} (c_s (-1+c_T^2) x_0 + 2 c_T (-1 + c_s c_T) y_0 ) (c_s (-1+c_T^2) x_0 + 2 c_T (1+c_s c_T) y_0 ) \nonumber \\
      & \times  (-2 y_0 + c_s^2 ( (-1+c_T^2 ) x_0 + 2 c_T^2 y_0 )   
       {\rm Plog}(2, e^{-x_0 - y_0}) + 16 c_s^2 \Big( \frac{(-1+c_T^2)^2 (-1+3 c_s^2 c_T^2)x_0^2}{c_T^2} \nonumber \\
       & + 12 (-1+c_T^2) (-1+c_s^2 c_T^2) x_0 y_0 + \frac{12 (-1+ c_s^2 c_T^2)^2 y_0^2 }{c_s^2}  \Big) {\rm Plog}(3,e^{-x_0-y_0})  + 192 (-1 + c_s^2 c_T^2) \nonumber \\
      & \times  (-2 y_0 + c_s^2 ((-1+c_T^2) x_0 + 2 c_T^2 y_0))  {\rm Plog}(4,e^{-x_0-y_0}) + 384 (-1 + c_s^2 c_T^2)^2 {\rm Plog}(5,e^{-x_0-y_0})  \Big] 
\end{align}
and
\begin{align}
     x_0 = \frac{c_T k}{T_{\rm today}} \left(\frac{g_{\star s}(T_{\rm max})}{g_{\star s}}\right)^{1/3}, \qquad y_0  = \frac{(1-c_T^2) }{2 (1 + c_s c_T)} x_0
\end{align}
for $y_0 \gg m$. 
As expected, the contributions vanish for $c_s = c_T=1$, as the interaction is forbidden by momentum and energy conservation. However, we still get a non-vanishing contribution for $c_T=1$ but $c_s < 1$, i.e.
\begin{align}
    \tilde \psi(x_0,y_0)\vert_{c_T=1} \simeq & \frac{(1-c_s^2)^2   }{8 c_s^6 \pi } x_0^6 \Big[  -   y_0^4   \log (1 - \cosh (x_0+y_0) + \sin(x_0+y_0) ) + 4   y_0^3      
       {\rm Plog}(2, e^{-x_0 - y_0})   \nonumber \\
      &  + 12  y_0^2   {\rm Plog}(3,e^{-x_0-y_0})  + 24  y_0  {\rm Plog}(4,e^{-x_0-y_0}) + 24  {\rm Plog}(5,e^{-x_0-y_0})  \Big]~.
\end{align}
The GW spectrum has a similar form as before and, therefore, we can get a sizeable spectrum for $c_T=1$ but $c_s \neq 1$, as it is indeed the higher derivative term which is mainly responsible for the production of the gravitational waves.

\bibliography{draft}

\end{document}